\documentclass{emulateapj}

\usepackage{graphicx}
\usepackage{wasysym}
\usepackage{natbib}
\usepackage[colorlinks,urlcolor=cyan,citecolor=blue,linkcolor=blue]{hyperref} 

\slugcomment{In press. Accepted to ApJL on 2013 August 20.}
\shortauthors{} 
\shorttitle{}
\begin{document}

\title{Inference of Inhomogeneous Clouds in an Exoplanet Atmosphere}

\author{
Brice-Olivier Demory\altaffilmark{1}, 
Julien de Wit\altaffilmark{1}, 
Nikole~Lewis\altaffilmark{1,10}, 
Jonathan~Fortney\altaffilmark{2}, 
Andras~Zsom\altaffilmark{1}, 
Sara~Seager\altaffilmark{1}, 
Heather~Knutson\altaffilmark{3},
Kevin~Heng\altaffilmark{4}, 
Nikku~Madhusudhan\altaffilmark{5},  
Michael~Gillon\altaffilmark{6},
Thomas~Barclay\altaffilmark{7},
Jean-Michel~Desert\altaffilmark{3},
Vivien~Parmentier\altaffilmark{8},
Nicolas~B.~Cowan\altaffilmark{9}
}

\altaffiltext{1}{Department of Earth, Atmospheric and Planetary Sciences, Massachusetts Institute of Technology, 77 Massachusetts Ave., Cambridge, MA 02139, USA. demory@mit.edu}
\altaffiltext{2}{Department of Astronomy and Astrophysics, University of California, Santa Cruz, CA 95064, USA}
\altaffiltext{3}{Division of Geological and Planetary Sciences, California Institute of Technology, Pasadena, CA 91125, USA}
\altaffiltext{4}{University of Bern, Center for Space and Habitability, Sidlerstrasse 5, CH-3012, Bern, Switzerland}
\altaffiltext{5}{Department of Physics \& Department of Astronomy, Yale University, New Haven, CT 06520, USA}
\altaffiltext{6}{Institut d'Astrophysique et de G\'eophysique, Universit\'e de Li\`ege, All\'ee du 6 Ao\^ut, 17, Bat. B5C, Li\`ege 1, Belgium.}
\altaffiltext{7}{NASA Ames Research Center, M/S 244-30, Moffett Field, CA 94035, USA}
\altaffiltext{8}{Laboratoire J.-L. Lagrange, UMR 7293, Universit\'e de Nice-Sophia Antipolis, CNRS, Observatoire de la C\^ote d'Azur B.P. 4229, 06304 Nice Cedex 4, France}
\altaffiltext{9}{Department of Physics and Astronomy, Northwestern University, 2145 Sheridan Rd., F165, Evanston, IL 60208, USA}
\altaffiltext{10}{Sagan Fellow}

\begin{abstract}
We present new visible and infrared observations of the hot Jupiter Kepler-7b to determine its atmospheric properties. Our analysis allows us to 1) refine Kepler-7b's relatively large geometric albedo of $Ag=0.35\pm0.02$, 2) place upper limits on Kepler-7b thermal emission that remains undetected in both {\it Spitzer} bandpasses and 3) report a westward shift in the {\it Kepler} optical phase curve. We argue that Kepler-7b's visible flux cannot be due to thermal emission or Rayleigh scattering from H$_2$ molecules. We therefore conclude that high altitude, optically reflective clouds located west from the substellar point are present in its atmosphere. We find that a silicate-based cloud composition is a possible candidate. Kepler-7b exhibits several properties that may make it particularly amenable to cloud formation in its upper atmosphere. These include a hot deep atmosphere that avoids a cloud cold trap, very low surface gravity to suppress cloud sedimentation, and a planetary equilibrium temperature in a range that allows for silicate clouds to potentially form in the visible atmosphere probed by {\it Kepler}. Our analysis does not only present evidence of optically thick clouds on Kepler-7b but also yields the first map of clouds in an exoplanet atmosphere.
\end{abstract}


\section{Introduction}
%%%%%%%

Clouds and hazes are ubiquitous in the Solar System's giant-planet and brown-dwarf atmospheres. In cloudy L-type brown dwarf atmospheres, the role of clouds has long been appreciated \citep[e.g.,][]{Ackerman:2001,Burrows:2001,Tsuji:2002,Kirkpatrick:2005,Witte:2009} and the observed spectra of such objects cannot be modeled correctly without clouds \citep{Cushing:2008}. It has been long suggested that clouds would also play a strong role in shaping the spectra of exoplanets in general \citep{Barman:2001,Marley:2013}, and hot Jupiters in particular \citep{Marley:1999,Seager:2000,Sudarsky:2000} before having been actually reported \citep{Evans:2013}.

Most hot Jupiters are ``dark'' at visible wavelengths \citep[e.g.,][]{Rowe:2008,Coughlin:2012a,Barclay:2012a} and only a handful exhibit appreciable geometric albedos. Caution is needed when interpreting hot-Jupiter geometric albedos, as for the most irradiated objects, a significant part of the planetary spectral energy distribution leaks into visible wavelengths, complicating the distinction between reflected light and thermal emission.

Kepler-7b \citep{Latham:2010} is a hot Jupiter orbiting a sub-giant G star in 4.89 days. Its relatively low mass $M_{p}=0.44\pm0.04$M$_{\rm Jup}$ and large radius $R_{p}=1.61\pm0.02$R$_{\rm Jup}$ result in a very low density $\rho_{p}=0.14$g cm$^{-3}$ \citep[][hereafter D11]{Demory:2011b}. Remarkably, Kepler-7b has a significant geometric albedo $A_{g}\sim0.35$ and exhibits a clear phase-curve modulation in the {\it Kepler} bandpass \citep[D11;][]{Kipping:2011d,Coughlin:2012a}. Kepler-7b's effective temperature places this hot Jupiter in an exceptionally rich region of condensation phase space. Because of the extreme difference between its equilibrium temperature and the brightness temperature as derived from its occultation in the {\it Kepler} bandpass, the origin of Kepler-7b's albedo has been attributed to the presence of a cloud or haze layer in its atmosphere or to Rayleigh scattering (D11). 

In this Letter, we use both optical phase-curve and infrared occultation data to determine the origin of Kepler-7b's visible flux. Section \ref{spitzer} presents the {\it Spitzer} observations and data analysis. Section \ref{kepler} describes our analysis of {\it Kepler} data employing three times more data than in D11. Section \ref{models} presents our discussion about the origin of flux observed in the {\it Kepler} bandpass.

\section{{\it Spitzer} 3.6- and 4.5-$\mu$m Photometry} 
\label{spitzer}
%%%%%%%

\subsection{Observations and Data Analysis}

We observed two occultations of Kepler-7b with {\it Spitzer} \citep{Werner:2004} in IRAC \citep{Fazio:2004a} 3.6-$\mu$m channel as well as two other in IRAC 4.5-$\mu$m channel between August and November 2011. All Astronomical Observation Requests (AORs) were obtained as part of program 80219 (PI H.Knutson). Datasets are $\sim$9-hour long and were obtained in full-array mode with an individual exposure time of 10.4s. A total of 2,440 frames was collected for each AOR. We perform a data reduction of all AORs similar to \citet{Demory:2011}, using as input the Basic Calibrated Data files produced by the {\it Spitzer} pipeline version 18.18.0. In a first step, we test twelve apertures ranging from 1.8 to 4.5 pixels and find the lowest RMS using 2.6 and 2.8-pixel apertures at 3.6 and 4.5 $\mu$m respectively. We obtain an RMS of 6380 and 6710 ppm for the two 4.5-$\mu$m AORs with a moderate contribution from correlated noise of less than 20\%. Our analysis of the 3.6-$\mu$m data resulted in significant correlated noise in both time-series ($>$40\%). Because of the long occultation duration of Kepler-7b (5.3 hours), the remaining out-of-transit photometry is small on each side of the eclipse, making the occultation parameters retrieval delicate in the presence of correlated noise. In a second step, we apply the noise-pixel variable aperture technique \citep{Lewis:2013} to all AORs. We find this method mitigates systematics found at 3.6$\mu$m. We report corresponding RMS of 4900 and 4750 ppm for both AORs in this channel using this technique, with a reduced correlated noise contribution of $\sim$15\%. We do not notice any improvement using noise-pixel aperture over the classical fixed-aperture photometry reduction at 4.5$\mu$m.

In order to model these data, we use the Markov Chain Monte Carlo (MCMC) implementation presented in \citet{Gillon:2012a}. We assume a circular orbit (D11), set the occultation depth as a jump parameter and impose priors on the orbital period $P$, transit duration $W$, time of minimum light $T_0$ and impact parameter $b=a \cos i/R_{\star}$ based on D11. For each MCMC fit (at 3.6 and 4.5$\mu$m), we run two chains of $10^5$ steps and assess their convergence using the statistical test from \citet{Gelman:1992}. 

We use the Bayesian Information Criterion (BIC) to select the optimal baseline model for our 4.5-$\mu$m observations. We find the most adequate model based on a classical second order $x$-$y$ polynomial \citep[][eq.~1]{Demory:2011} to correct the ``pixel-phase'' effect, added to a time-dependent linear trend.  The baseline model for our 3.6-$\mu$m data consists of the noise-pixel parameter alone. We discard the first $\sim$25-35 minutes of all AORs that are affected by a noticeable detector ramp and/or increased noise, already noticed in warm-Spitzer photometry \citep[e.g.,][]{Deming:2011a}. Our {\it Spitzer}/IRAC raw lightcurves are shown on Fig.\ref{fig_spitzer} (left).

\subsection{The Thermal Emission of Kepler-7b}

\begin{figure*}
\centering
   \epsscale{1.0}\plottwo{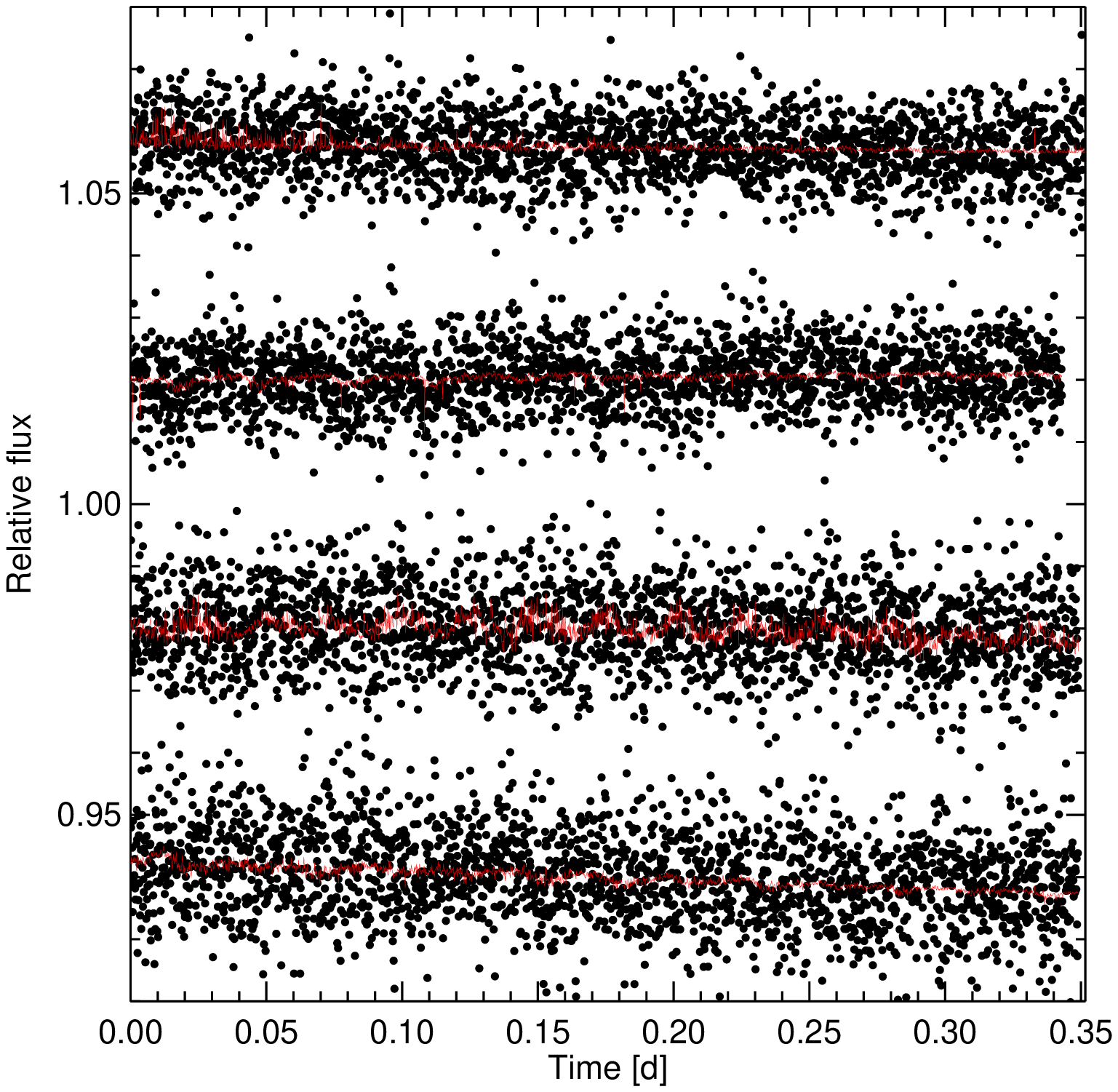}{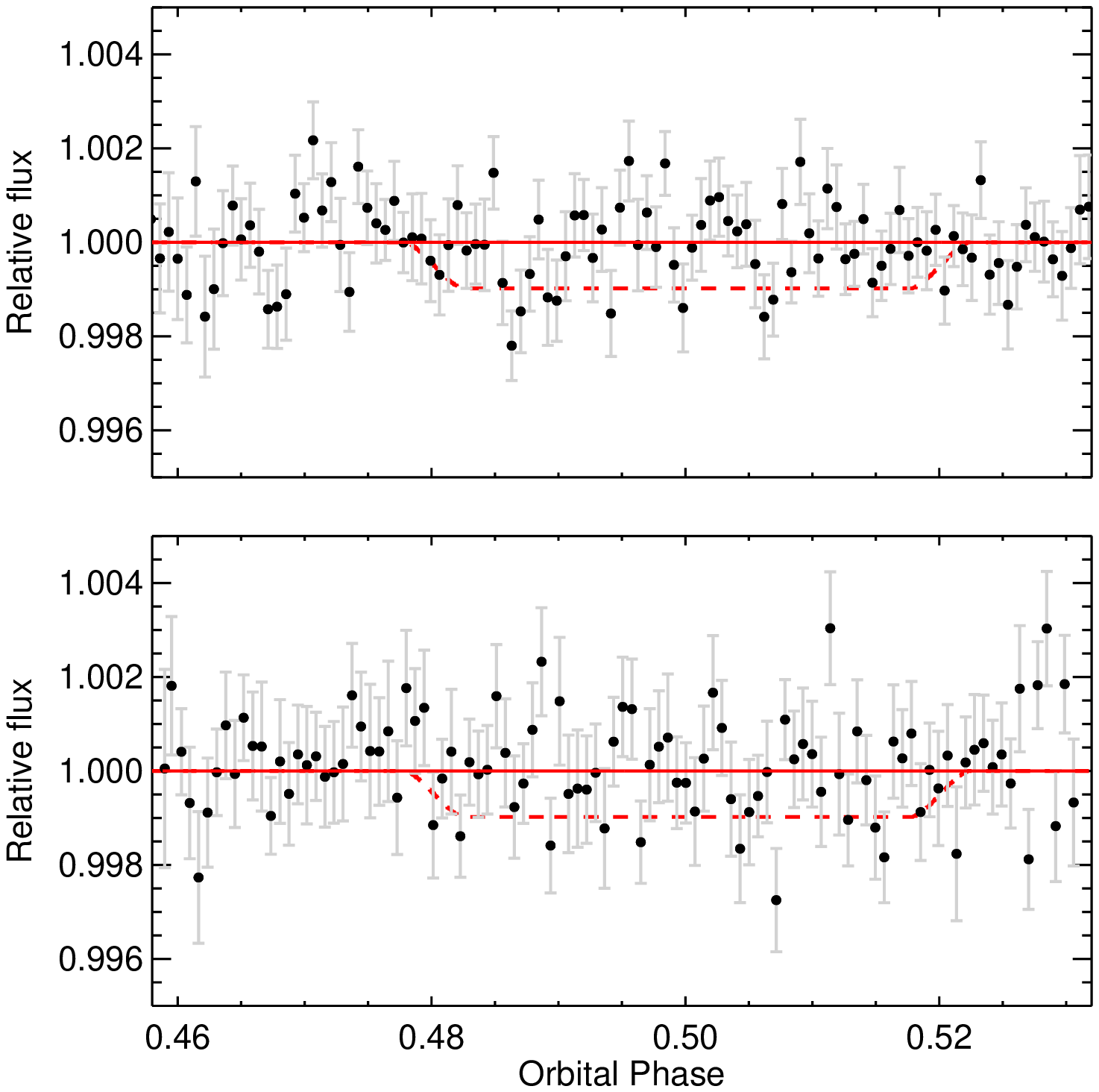}
  \caption{{\bf {\it Spitzer} 3.6 and 4.5-$\mu$m occultation photometry of Kepler-7b.} {\it Left:} raw photometry of the four AORs with the best-fit model superimposed (see Sect.\ref{spitzer}). The lightcurves are shifted on the vertical axis for clarity. The two IRAC 3.6-$\mu$m lightcurves are at the top and the two 4.5-$\mu$m lightcurves at the bottom.    {\it Right:} Phase-folded occultations divided by the best-fit model. The IRAC 3.6-$\mu$m lightcurve is shown at the top and the 4.5-$\mu$m at the bottom. Data are binned per 5 minutes. For illustration purposes we depict a 1-mmag occultation in red dash-line, the best-fit model for the two channels being a null occultation.}  \label{fig_spitzer}
\end{figure*}

We repeat the same MCMC fits for both channels setting the occultation depth to zero, to compare the BIC between a model that includes the occultation and a model that does not. The MCMC fits including the occultation model yield an occultation depth of 164$\pm$150 ppm at 3.6$\mu$m and 367$\pm$221 ppm at 4.5$\mu$m. We compare the BIC of these runs to the MCMC fits that do not include the occultation model. The odds ratio between both models is $\sim$180 and $\sim$100 in favor of the model without occultation at 3.6 and 4.5$\mu$m respectively. Based on our dataset, the occultation is detected in none of the channels. We derive corresponding 3-$\sigma$ upper limits of 615 and 1010 ppm at 3.6 and 4.5$\mu$m. We employ a \textsc{PHOENIX} \citep{Hauschildt:1999} model of Kepler-7 using the D11 stellar parameters to convert these occultation depth upper-limits into brightness temperatures. We find these 3-$\sigma$ upper-limits to be 1700 and 1840K at 3.6 and 4.5$\mu$m. Our final phase-folded occultation lightcurves are shown on Fig.\ref{fig_spitzer} (right).

\section{{\it Kepler} Observations and Data Analysis} 
\label{kepler}
%%%%%%%

\subsection{Data Reduction}

We base our analysis on {\it Kepler} \citep{Batalha:2013} quarters 1-14 long-cadence simple aperture photometry \citep{Jenkins:2010a} that span more than 1,200 days of quasi-continuous observations, which is three times more data than in D11. We mitigate instrumental systematics by fitting the first four cotrending basis vectors (CBV) to each quarter using the PyKE software \citep{Still:2012}. We find that outliers represent less than $\sim$0.5\% of the dataset. We then normalize each quarter to the median. In total, 56,000 datapoints are collected. We employ our MCMC framework presented in Sect.\ref{spitzer} to account for photometric trends longer than twice the planetary orbital period by fitting a second-order polynomial to the out-of-eclipse data.

We then evaluate the contribution from correlated noise on timescales corresponding to the orbital period. We cut the whole data into 5-day duration segments and compute a scaling factor $\beta$ based on the standard deviation of the binned residuals for each light curve using different time-bins \citep{Gillon:2010a}. We keep the largest $\beta$ value as a criterion to discard data segments affected by significant correlated noise. We obtain a mean $\beta=1.19$ over the whole data set and discard those with threshold $\beta>2.1$, which account for $\sim$5\% of the complete dataset. All data discarded affect predominantly quarters 12-14, when increased solar activity and coronal mass ejections resulted in a decrease of {\it Kepler}'s pointing accuracy and thus an increase in systematic noise. We finally note that in contrary to pre-whitening techniques (as employed in D11), the data-reduction method presented here preserves all phase-curve properties.

\subsection{Robustness of the planetary phase-curve signal}
 
To assess the robustness of the phase-curve properties, we repeat the analysis presented above several times, by increasing the number of CBV components up to 8, by decreasing the threshold $\beta$ values and by using linear or third-order polynomials to account for the long-term trends. We find the phase amplitude, peak-offset and occultation depth values to remain consistent within 1-$\sigma$ uncertainties (see Sect.\ref{phase-analysis}). The phase-curve signal is therefore not due to (nor affected by) the detrending. Two of us (BOD, TB) performed independent analyses of the dataset and obtained results in excellent agreement.

\begin{figure*}
\centering
   \epsscale{0.8}\plotone{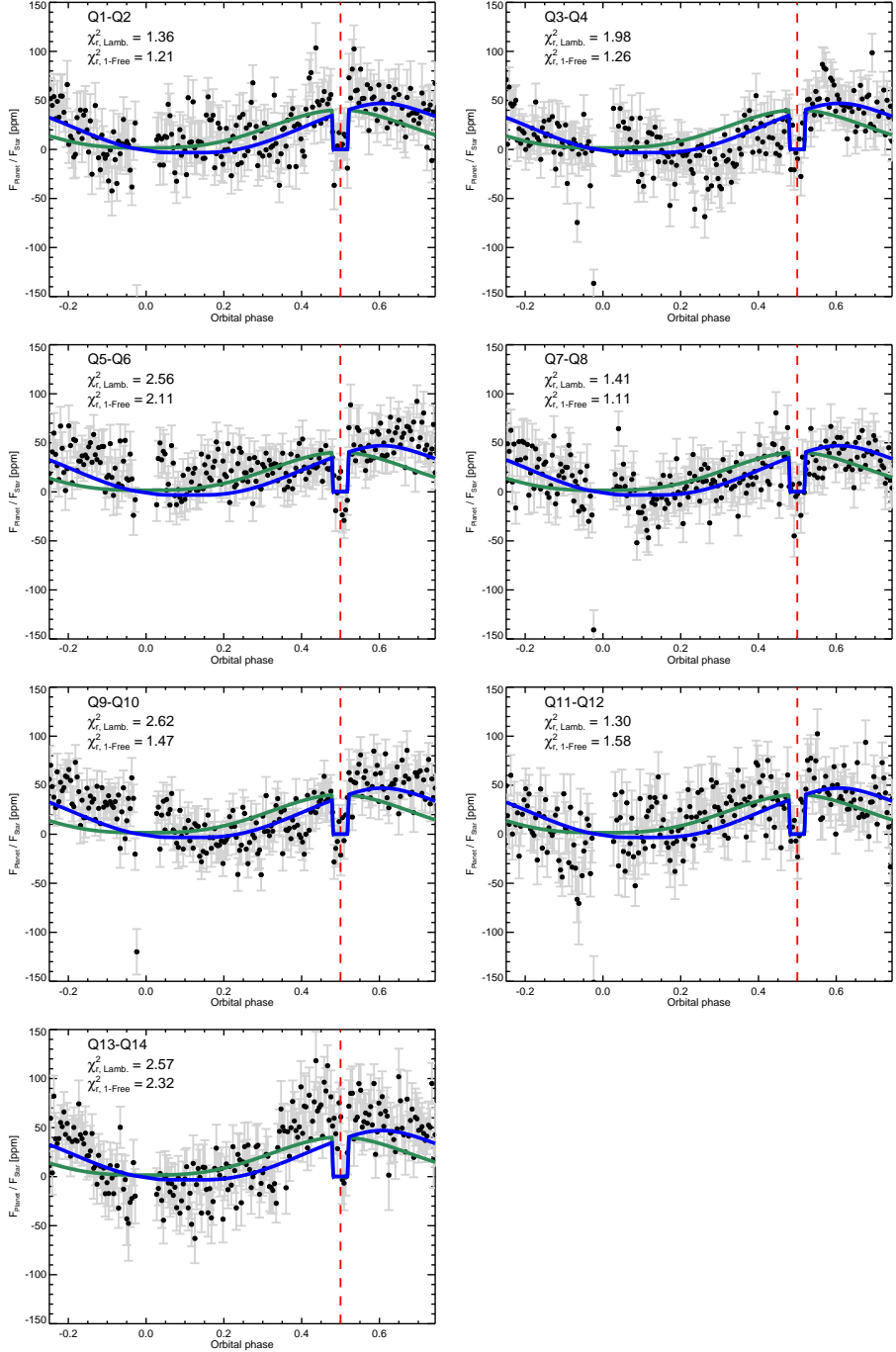}
  \caption{{\bf Matrix of Kepler-7b phase curves based on pairs of {\it Kepler} quarters.} Data are binned per 5 minutes. The symmetric Lambertian sphere (green) and asymmetric 1-free-band model (blue) models are superimposed, along with the corresponding $\chi^{2}_{r}$ values (See Sect.\ref{phase-analysis}). The occultation's phase is indicated in red. The asymmetric model is preferred for all quarter pairs, excepted Q11-Q12.}  \label{fig_kep_q}
\end{figure*}

\begin{figure*}
\centering
   \epsscale{1.0}\plotone{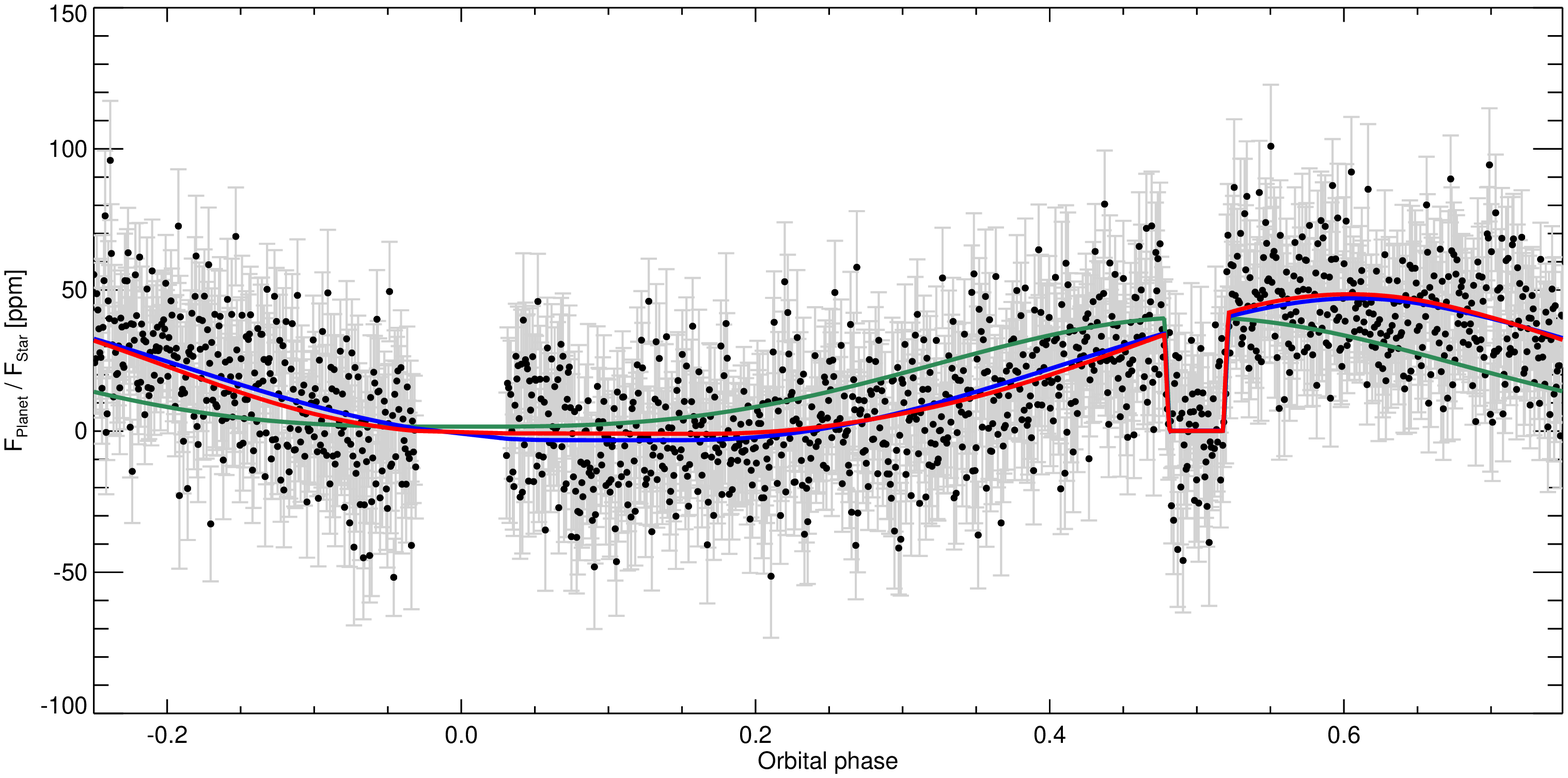}
  \caption{{\bf Phase curve of Kepler-7b based on {\it Kepler} Q1-Q14 data.} Data are binned per 5 minutes. The Lambertian sphere (green), 1-free-band (blue) and 3-fixed-band (red) best-fit models (see Sect.\ref{phase-analysis}) are superimposed. }  \label{fig_kep}
\end{figure*}

Figures \ref{fig_kep_q} and \ref{fig_kep} demonstrate the stability of the phase-curve signal across Q1-Q14. This would not be the case if the phase curve was of instrumental origin as while {\it Kepler} systematics can be consistent in amplitude across quarters, they are definitely not consistent in phase \citep[e.g.,][]{Kinemuchi:2012}. Any signal due to {\it Kepler} systematics would thus average out across quarters. This strongly favors the phase-curve being of astrophysical origin.

We search for all frequencies in the dataset to assess any risk of contamination of the planetary phase curve. To quantify how frequencies and amplitudes evolve with time, we perform a wavelet transform analysis using the weighted wavelet Z-transform algorithm developed by \citet{Foster:1996}. We do not detect any clear signature, apart from the planet orbital signal. Kepler-7 is intrinsically quiet and any stellar activity remains nominal over Q1-Q14 observations, with no quarter-dependent fluctuations. We notice a barely detectable periodicity at $\sim$16.7 days that could correspond to the rotational period of the star, which translates to an equatorial velocity of $V_{eq}\sim6$ km/s assuming $R_{\star}=2.02R_{\odot}$ (D11). This is broadly consistent with Kepler-7's stellar projected rotation $v{\rm \sin}i=4.2$ km s$^{-1}$ \citep{Latham:2010}.

The host star is unlikely to contaminate our phase-curve for several reasons. As we phase-fold data over more 3.5 years, only stellar variability exactly phased on the planetary orbital period (or a multiple) and consistent over the duration of the observations could affect the phase-curve shape. First, the stellar rotational velocity suggests that the star is not tidally locked to the planet, as the planetary orbital period is only $\sim$4.89 days. The stellar rotation and planetary orbital periods are different by a non-integer factor of $\sim$3.4. Second, stellar pulsations with a period of $\sim$5 days are unlikely for a sub-giant star and would have been visible in the data. Third, as we do not clearly detect stellar variability in the photometry, only small starspots could be present, but those starspots would have a short lifetime \citep[e.g.,][]{Strassmeier:2009}. Even in the case of starspots that are stable over more than three years, differential rotation would cause distortions in the lightcurve across quarters that are not observed (Fig.\ref{fig_kep_q}). Furthermore, spots or group of spots do not usually produce sinusoidal lightcurves but rather sequences of flat and V-shaped lightcurves \citep[e.g.,][]{Harrison:2012}. Finally, we do not detect interactions between the star and the planet in the form of ellipsoidal or beaming components in the phase curve.

We finally take into account a faint stellar companion located 1.9''-East of Kepler-7 with a $\Delta$mag$=4.0$ both in $J$ and $Ks$ bands \citep{Adams:2012}. These flux ratios suggest a similar spectral type and discard the possibility of a cool star. In order to detect a significant contamination from the companion star with a period commensurate with Kepler-7b's orbital period, we split the full dataset in segments of duration equal to a quarter. Each quarter has a specific aperture with a different contribution from the fainter companion star. The reported consistency at the 1-$\sigma$ level of the phase curve properties (amplitude, phase-peak offset) across quarters suggest a negligible contamination from the stellar neighbor. 

We therefore conclude that the phase curve is of planetary origin.

%%%%%

\subsection{Phase Curve Analysis}
\label{phase-analysis}

Kepler-7b's phase curve deviates from a pure Lambert-law phase-dependent behavior \citep[e.g.,][]{Sobolev:1975} expected for isotropic scattering alone (Fig.\ref{fig_kep}, green). The main feature of Kepler-7b's phase curve is a delay of $13\pm3.5$ hours of the phase-curve's peak from the occultation center. This delay implies that the hemisphere-integrated flux is maximum to the west of Kepler-7b's substellar point. We further measure a phase-curve amplitude of $50\pm2$ ppm and an occultation depth of $48\pm3$ ppm, corresponding to a geometric albedo $A_g=0.35\pm0.02$.  This occultation depth translates to a brightness temperature of $2645^{+20}_{-30}$K in the {\it Kepler} bandpass, which is 1000K and 800K larger than the infrared brightness temperatures upper limits measured at 3.6 and 4.5$\mu$m respectively (see Sect.\ref{spitzer}). We found our phase-curve amplitude and occultation depth to be in  agreement with previous analyses \citep[D11;][]{Kipping:2011d,Coughlin:2012a}.

The key features of Kepler-7b's phase-curve translate directly into constraints on maps \citep{Cowan:2008} assuming a tidally-locked planet on a circular orbit. A planetary phase-curve $\frac{F_p}{F_{\star}}$ measures the planetary hemisphere-averaged relative brightness $\frac{<I_p>}{<I_{\star}>}$ as follows:
\begin{equation}
\frac{F_p}{F_{\star}}(\alpha)=\frac{<I_p>(\alpha)}{<I_{\star}>}\left(\frac{R_p}{R_{\star}}\right)^2  
\label{eq1}
\end{equation}
where $\alpha$ is the orbital phase.

We first notice that Kepler-7b's planetary flux contribution starts from phase $0.18\pm0.03$, when the meridian centered $25\pm12^{\circ}$ East of the substellar point appears. Second, the phase-curve's maximum is located at phase $0.61\pm0.03$, implying that the brightest hemisphere is centered on the meridian located $41\pm12^{\circ}$ West of the substellar point. Third, the planetary flux contribution vanishes around the transit, implying that the ``bright'' area extends up to the western terminator, while its extension to the East of the substellar point is nominal. We finally note that the phase-curve's amplitude of $50\pm2$ ppm converts into an hemisphere-averaged relative brightness $74\pm2\times10^{-4}$ (eq.\ref{eq1}).

\begin{figure*}
\centering
  \epsscale{1.0}\plotone{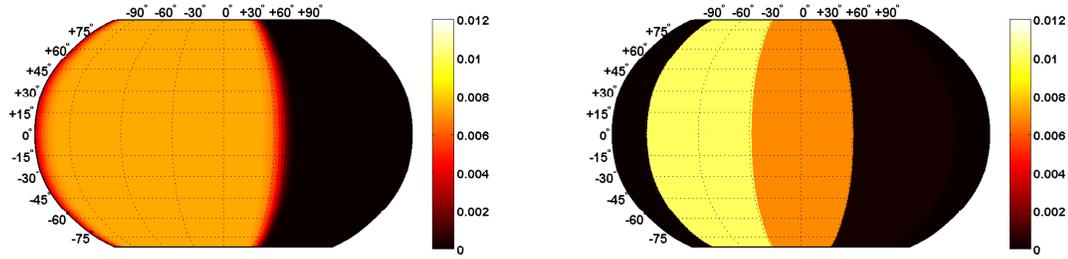}
    \caption{{\bf Longitudinal brightness maps of Kepler-7b.} Kepler-7b's longitudinal brightness distributions $\frac{I_p}{I_{\star}}$ as retrieved in {\it Kepler}'s bandpass using the 1-free-band model (left) and the 3-fixed-band model (right) detailed in Sect.\ref{phase-analysis}).}  \label{fig_kep2}
\end{figure*}

We longitudinally map Kepler-7b using the MCMC implementation presented in \citet{de-Wit:2012a}. This method has been developed to map exoplanets and to mitigate the degeneracy between the planetary brightness distribution and the system parameters. We use two model families similar to the ``beach-ball models'' introduced by \citet{Cowan:2009b}: one using $n$ longitudinal bands with fixed positions on the dayside and another using longitudinal bands whose positions and widths are jump parameters in the MCMC fit. We choose the two simplest models from these families: a 3-fixed-band model and 1-free-band model so as to extract Kepler-7b's longitudinal dependence of the dayside brightness as well as the extent of the ``bright'' area. For both models, we compute each band's amplitude from their simulated lightcurve by using a perturbed singular value decomposition method. The corresponding median brightness maps are shown on Fig.\ref{fig_kep2}. The 1-free-band model (Fig.\ref{fig_kep}, blue) finds a uniformly-bright longitudinal area extending from $105\pm12^{\circ}$ West to $30\pm12^{\circ}$ East with a relative brightness $78\pm4\times10^{-4}$ (Fig.\ref{fig_kep2}, left). The 3-fixed-band model (Fig.\ref{fig_kep}, red) finds bands of relative brightness decreasing from the West to the East with the following values: 100 to 68 and 3$\pm6\times10^{-4}$ (Fig.\ref{fig_kep2}, right). We finally note that the 1-free-band model finds a bright sector extending to the night side, due to the sharp flux increase observed around transit (Fig.\ref{fig_kep}).

\section{The Origin of Kepler-7b's Visible Flux}
\label{models}
%%%%%%%

The combined information from the {\it Spitzer} and {\it Kepler} observations of Kepler-7b strongly favor the conclusion that the planetary phase-dependent flux variations seen in the {\it Kepler} light curve are the result of scattered light from optically thick clouds, whose properties change as a function of longitude.

The lack of significant thermal emission from Kepler-7b in the {\it Spitzer} 3.6 and 4.5-$\mu$m bandpasses supports the fact that Kepler-7b's visible light curve is driven by reflected light. Kepler-7\,b's phase curve exhibits a westward asymmetry suggesting, if of thermal origin, a temperature structure that does not follow the expected temperature structure for tidally-locked hot Jupiters, which would yield an eastward shift. This eastward shift is consistently produced from a range of general circulation models for tidally-locked hot-Jupiters forced using various methods, including Newtonian cooling \citep[e.g.][]{Cooper:2005,Showman:2008,Dobbs-Dixon:2010,Rauscher:2010, Heng:2011}, dual-band radiative transfer \citep[e.g.][]{Heng:2011a,Rauscher:2012} or multi-wavelength radiative transfer \citep[e.g.][]{Showman:2009}. Combining these results with the analytical theory of \citet{Showman:2011a} suggests that thermal phase-curve eastward shifts are robust outcomes of the hot-Jupiter circulation regime. As we do not detect thermal flux from Kepler-7b with \emph{Spitzer}, the most likely conclusion is that the westward shift in the visible phase-curve is indicative of a variation in the cloud properties (cloud coverage, optical depth, particle size distribution, vertical extent, composition, etc.) as a function of longitude, governed by the planet's wind and thermal patterns.

\begin{figure*}
\centering
  \plottwo{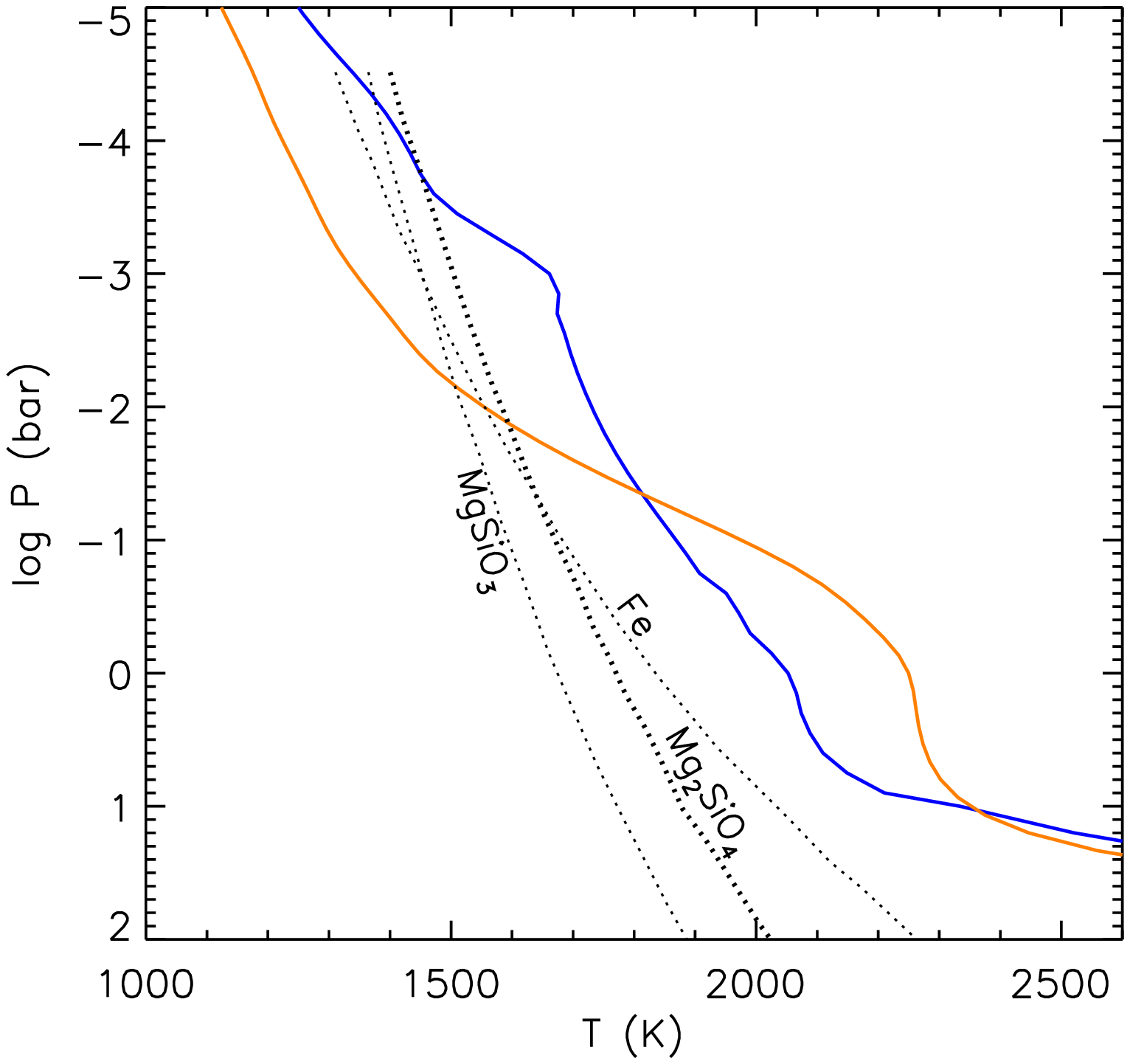}{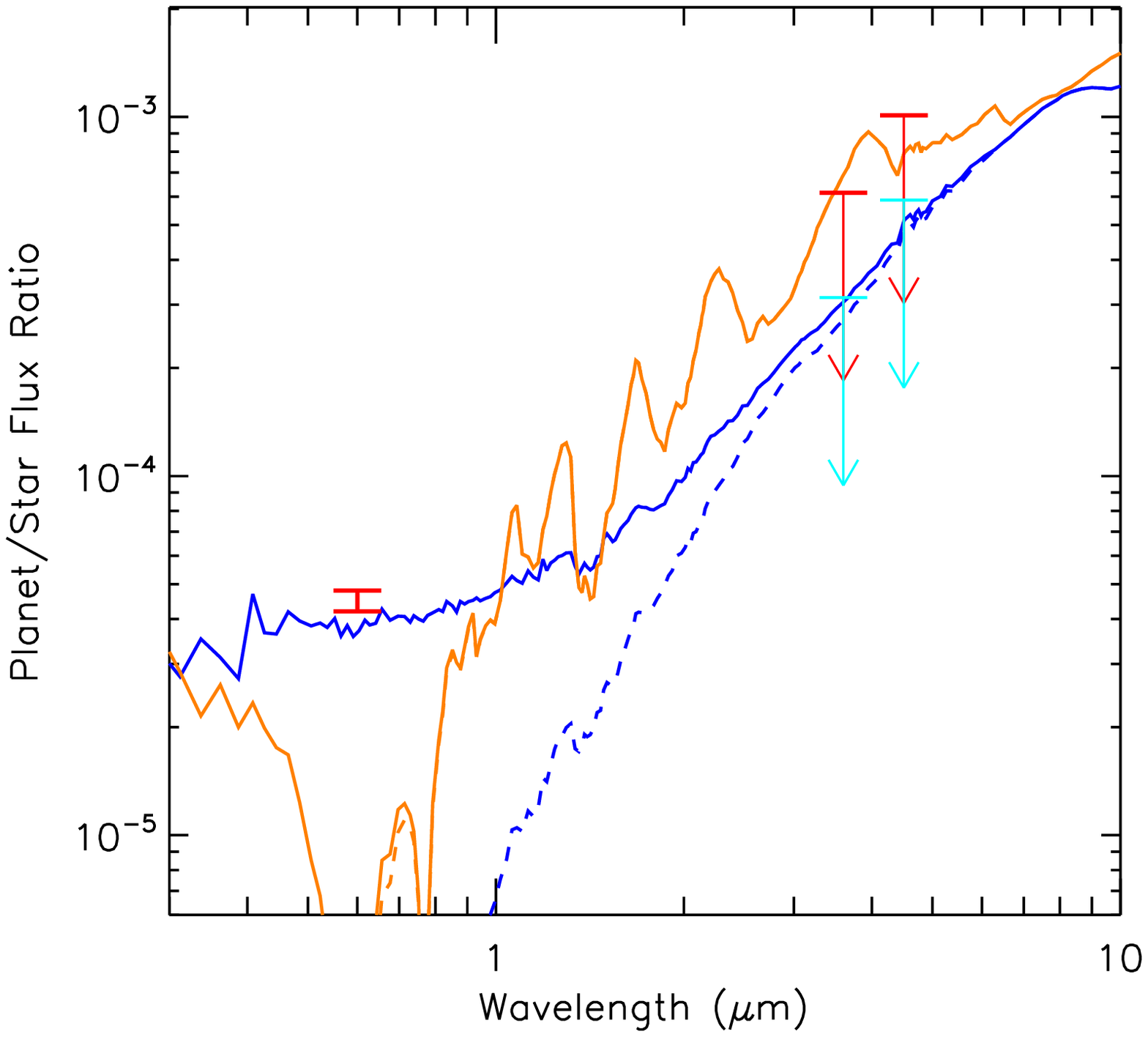}
  \caption{{\bf One-dimensional models of the dayside temperature structure and flux ratios of Kepler-7b.}  {\it Left:} Condensation curves for several species are also shown, although only Mg$_2$SiO$_4$ is used in the calculations.  The model in orange is cloud-free, while the model in blue includes cloud opacity. {\it Right:} The cloud-free model is dark in the optical and emits more flux in the mid-infrared IRAC bands.  Dashed curves are the thermal emission component and solid curves are the total flux.  The cloudy model is brighter in the optical, owing to scattered light, with suppression of mid-infrared flux.  The optical detection in the \emph{Kepler} band is shown in red, along with the \emph{Spitzer} 1-$\sigma$ (cyan) and 3-$\sigma$ (red) upper limits.} \label{jfpt}
\end{figure*}

We use the methods of \citet{Fortney:2005,Fortney:2008} to compute Kepler-7b's one-dimensional (1D) temperature structure and emission spectrum (Fig.\ref{jfpt}). The orange model is cloud-free.  The blue model uses the cloud model of \citet{Ackerman:2001} to calculate the vertical distribution and optical depths of Mg$_2$SiO$_4$ clouds. Both models assume modest redistribution of energy, with the assumption that 1/4 of the incident energy is lost to the un-modeled night side. The particle size distribution in the cloud is assumed to be log-normal with a mode of 0.5$\mu$m at all heights. A low sedimentation efficiency free parameter ($f_{sed}$) of 0.1 is used, which suppresses sedimentation.

It is clear that the cloudy model (blue) provides a much better fit to the combined occultation measurements from {\it Spitzer} and {\it Kepler}.  The clouds dramatically enhance the flux in the optical, increase the model Bond albedo, and suppress emission in the infrared (Fig.\ref{jfpt}, right). We note that many other combinations of cloud and thermal properties might also provide an adequate match to the data. However, we exclude Rayleigh scattering from H$_2$ molecules and homogeneous cloud structures as possible sources of visible phase-curve signatures, which would both result in a symmetric phase curve.

Kepler-7b may be relatively more likely to show the effects of cloud opacity than other hot Jupiters. The planet's incident flux level is such that model profiles cross silicate condensation curves in the upper, observable atmosphere, making these clouds a possible explanation.  The same would not be true for warmer planets (where temperatures would be too hot for dayside clouds) or for cooler planets (where silicates would only be present in the deep unobservable atmosphere).  Furthermore, the planet's very low surface gravity may play an important role in hampering sedimentation of particles out the atmosphere. Finally, the planet's large radius implies a relatively high specific entropy adiabatic in the interior, and a correspondingly warm adiabat in the deep atmosphere at tens of bars.  This removes the possibility of silicate clouds condensing at pressures of 100-1000 bars, as may happen in other hot Jupiters. 

Our results suggest that one broad-band visible phase curve is probably insufficient to constrain the cloud properties. The problem might remain degenerate until more observations (such as narrow-band optical phase curves and polarimetry) become available.
In the near future it is likely that similar brightness maps of other {\it Kepler} planets will emerge, thereby providing an invaluable means to improve our understanding of cloud formation in exoplanet atmospheres.

\acknowledgments
We thank G. Basri and both anonymous referees for helpful comments that improved the paper. This work is based in part on observations made with the \textit{Spitzer Space Telescope}, which is operated by the JPL, Caltech under a contract with NASA. A. Zsom was supported by the German Science Foundation (DFG) under grant ZS107/2-1. This work was performed in part under contract with the California Institute of Technology funded by NASA through the Sagan Fellowship Program executed by the NASA Exoplanet Science Institute. J. de Wit acknowledges support from the Belgian American Educational Foundation and Wallonie-Bruxelles International. 

{\it Facilities:}\facility{{\it Kepler, Spitzer}}

%\bibliography{apj-jour,kepler-7b}

\end{document}